\documentclass[twocolumn]{aastex62}

\usepackage{natbib}
\usepackage{amsmath}
\usepackage{braket}
\begin{document}
\def\vec#1{\mbox{\boldmath $#1$}}
\def\revise#1{\textcolor{red}{{\bf #1}}}
\newcommand{\average}[1]{\ensuremath{\langle#1\rangle} }
\renewcommand{\floatpagefraction}{0.99}
\renewcommand{\textfraction}{.10}

\title{\large{\bf{Anisotropic magnetohydrodynamic turbulence driven by parametric decay instability: \\
the onset of phase mixing and Alfv\'en wave turbulence}}}

	\author[0000-0002-7136-8190]{Munehito Shoda}
	\affiliation{Department of Earth and Planetary Science, The University of Tokyo, Hongo, Bunkyo-ku, Tokyo, 113-0033, Japan }
	\author[0000-0001-5457-4999]{Takaaki Yokoyama}
	\affiliation{Department of Earth and Planetary Science, The University of Tokyo, Hongo, Bunkyo-ku, Tokyo, 113-0033, Japan }
	
	\correspondingauthor{Munehito Shoda}
	\email{shoda@eps.s.u-tokyo.ac.jp}

\begin{abstract}

	We conduct a three-dimensional magnetohydrodynamic (MHD) simulation of the parametric decay instability of Alfv\'en waves and resultant compressible MHD turbulence, which is likely to develop in the solar wind acceleration region.
	Because of the presence of the mean magnetic field, the nonlinear stage is characterized by filament-like structuring and anisotropic cascading.
	By calculating the timescales of phase mixing and the evolution of Alfv\'en wave turbulence, we have found that the early nonlinear stage is dominated by phase mixing, 
	while the later phase is dominated by imbalanced Alfv\'en wave turbulence.
	Our results indicate that the regions in the solar atmosphere with large density fluctuation, such as the coronal bottom and wind acceleration region, are heated by phase-mixed Alfv\'en waves, 
	while the other regions are heated by Alfv\'en wave turbulence.
		
\end{abstract}


\section{Introduction} \label{sec:introduction}

	Since the discovery of the high-temperature solar atmosphere and supersonic outflow from the Sun,
	coronal heating and solar wind acceleration have become one of the most important research subjects in astrophysics.
	{\it Parker Solar Probe} \citep{Fox0016} is going to be launched to resolve these long-standing problems in solar physics.
		
	In the open field regions, the wave/turbulence model is promising as the heating mechanism \citep{Cranm12b}.
	Several photospheric observations have shown that transverse waves have sufficient energy flux at the photosphere \citep{Fujim09,Chitt12},
	and a certain portion of these waves propagates through the chromosphere into the corona to power the solar wind \citep{DePon07a,McInt11}.
	Non-thermal coronal line broadening \citep{Baner09,Hahn013} also indicates sufficiently strong Alfv\'en waves in the corona.
	
	While the global energetics of the corona and solar wind is clarified by wave observations, the energy cascading mechanism is still under investigation.
	We note that, because there is a large gap between the energy-containing scale and the dissipation scale of Alfv\'en waves in the corona and solar wind,
	the cascading rate gives the approximate plasma heating rate.
	There are mainly three promising cascading mechanisms: Alfv\'en wave turbulence (AWT) \citep{Dobro80,Perez13}, phase mixing (PM) \citep{Heyva83,Magya17}, and parametric decay instability (PDI) \citep{Suzuk05,Tener13}.
	
	AWT is triggered by counter-propagating Alfv\'en waves \citep{Kraic65,Goldr95}.
	In the corona and solar wind, partial reflection due to the field-aligned gradient of Alfv\'en speed \citep{Velli93,Verdi07} can drive AWT.
	Some theoretical models reproduce the corona and solar wind self-consistently based on the AWT heating scenario \citep{Cranm07,Verdi10,Chand11}.
	However, recent three-dimensional simulations indicate that the heating by AWT is insufficient \citep{Perez13,Balle17}.
	
	PM works when the Alfv\'en speed is inhomogeneous in the direction perpendicular to the mean magnetic field \citep{Heyva83}.
	Since the horizontal density variation is observed in the corona \citep{Raymo14}, PM plays a role in the coronal heating.
	In fact, a similar process called resonant absorption is observed to work in the solar atmosphere \citep{Okamo15}.
	A recent magnetohydrodynamic simulation has shown that PM can generate a turbulent structure, and this process is called generalized phase mixing \citep{Magya17}.
			
	PDI is an instability of the Alfv\'en wave \citep{Sagde69,Golds78},
	in which an Alfv\'en wave decays into a forward slow-mode wave and backward Alfv\'en wave in a low-beta regime.
	PDI can be a source of coronal heating because slow shocks are generated, heating up the corona and solar wind \citep{Suzuk05,Shoda18a}.
	Several processes in the solar wind are possibly attributable to PDI, such as the cross helicity evolution \citep{Malar00,Shoda16}, density fluctuation \citep{Bowen18,Shoda18c}, and power spectrum formation \citep{Chand18}.
	Both numerical simulation \citep{Suzuk05,Shoda18a} and radio-wave observation \citep{Miyam14} show large density fluctuations in the wind acceleration region, possibly generated by PDI.
	These results indicate that the plasmas in that region are in a state of PDI-driven turbulence.
		
	In this study, we conduct a simulation of PDI-driven turbulence to clarify which of the process mentioned above is the most responsible for energy cascading (and dissipation).
	Multi-dimensional simulations of PDI-driven turbulence have already been performed, focusing mainly on the spectral property \citep{Ghosh94a} or the dependence on dimensions \citep{DelZa01},
	although no simulations have ever been performed focusing on the cascading process.
	We perform a detailed analysis to reveal the dominant process in each phase of instability and turbulence.

\section{Method}  \label{sec:method}

\begin{figure*}[th]
	\begin{center}
	\includegraphics[width=140mm]{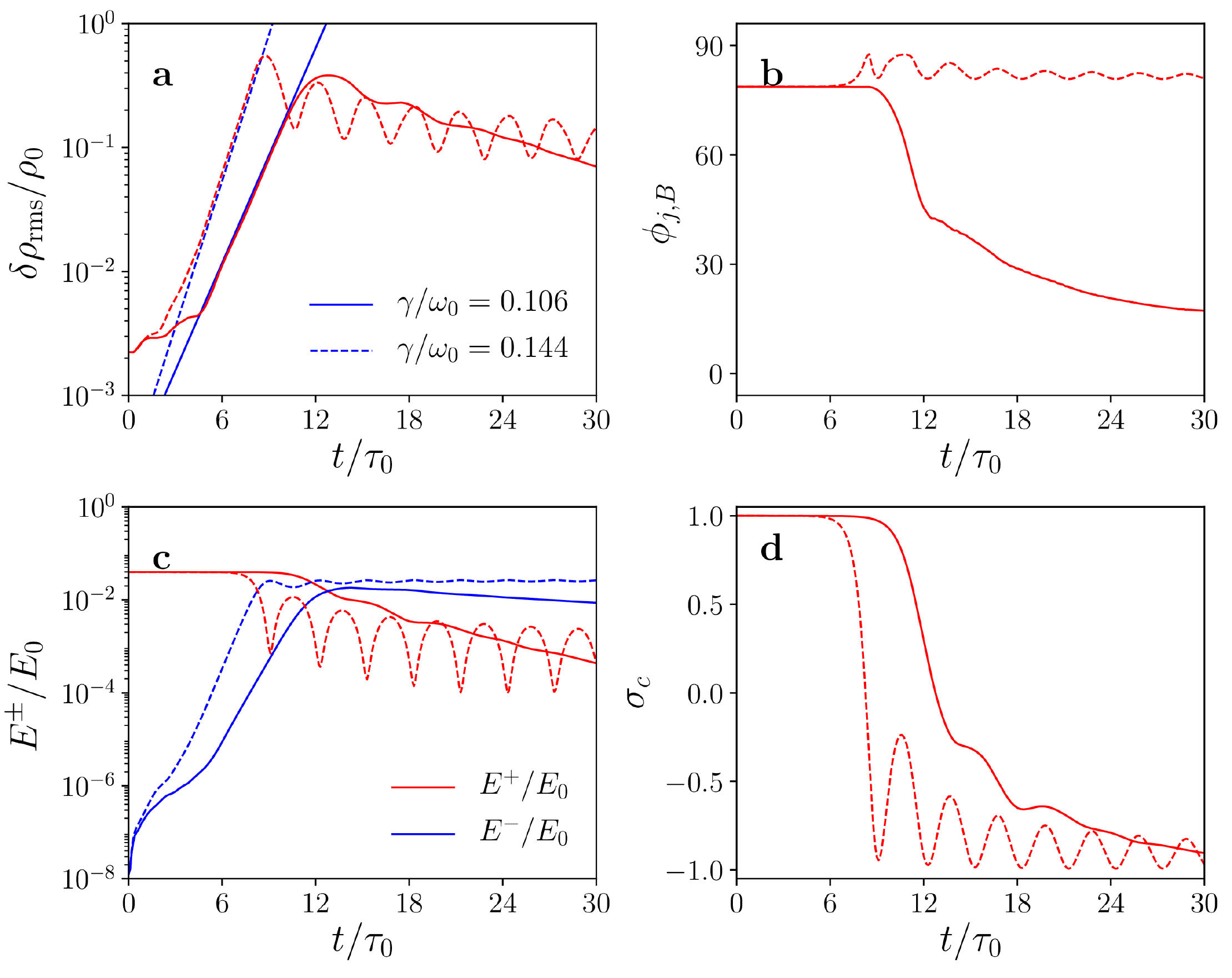} 
	\end{center}
	\vspace{-2em}
	\caption{
			Time evolution of the volume-averaged variables in three-dimensional setting (solid lines) and one-dimensional setting (dashed lines).
			The horizontal axis (time) is normalized by the period of the parent wave $\tau_0$.
			Panels indicate the a: normalized root-mean-square density fluctuation $\delta \rho_{\rm rms} / \rho_0$, b: angle between current density and magnetic field $\phi_{j,B}$, c: normalized Els\"asser energies (red: $E^+/E_0$, blue: $E^-/E_0$), and d: normalized cross helicity $\sigma_c$, respectively.
			The blue lines in Panel a are the exponential fitting  $\delta \rho_{\rm rms} \propto \exp \left( \gamma t \right)$ in the linear growth stage.
						}
	\vspace{2em}
	\label{fig:fundamental}
	\end{figure*}

	\begin{figure*}[th]
	\begin{center}
	\includegraphics[width=185mm]{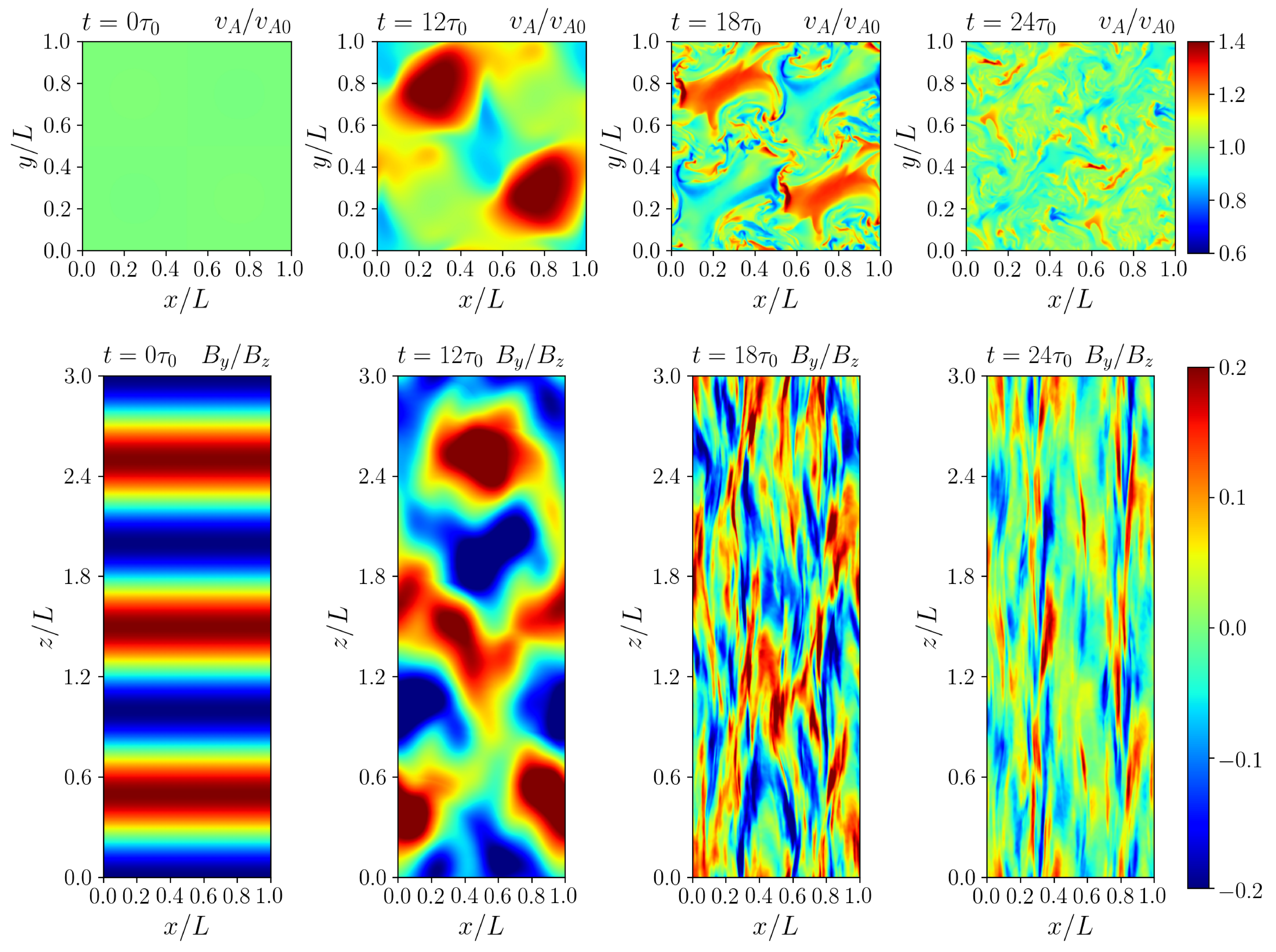} 
	\end{center}
	\vspace{-1em}
	\caption{
		Snapshots of the normalized Alfv\'en speed $v_{A} / v_{A0}$ on the $xy$ plane ($z=0$, upper panels) and normalized amplitude of Alfv\'en wave $B_y / B_z$ on the $xz$ plane ($y=0$, lower panels).
          	From left to right, the snapshots shown are $t=0$ (initial condition), $t = 12/\tau_0$ (saturation phase), $t=18 \tau_0$ (early nonlinear phase), and $t=24 \tau_0$ (fully nonlinear phase). \\
		(An animation of this figure is available.)
						}
	\vspace{1em}
	\label{fig:caby}
	\end{figure*}
	
	\begin{figure*}[th]
	\begin{center}
	\includegraphics[width=185mm]{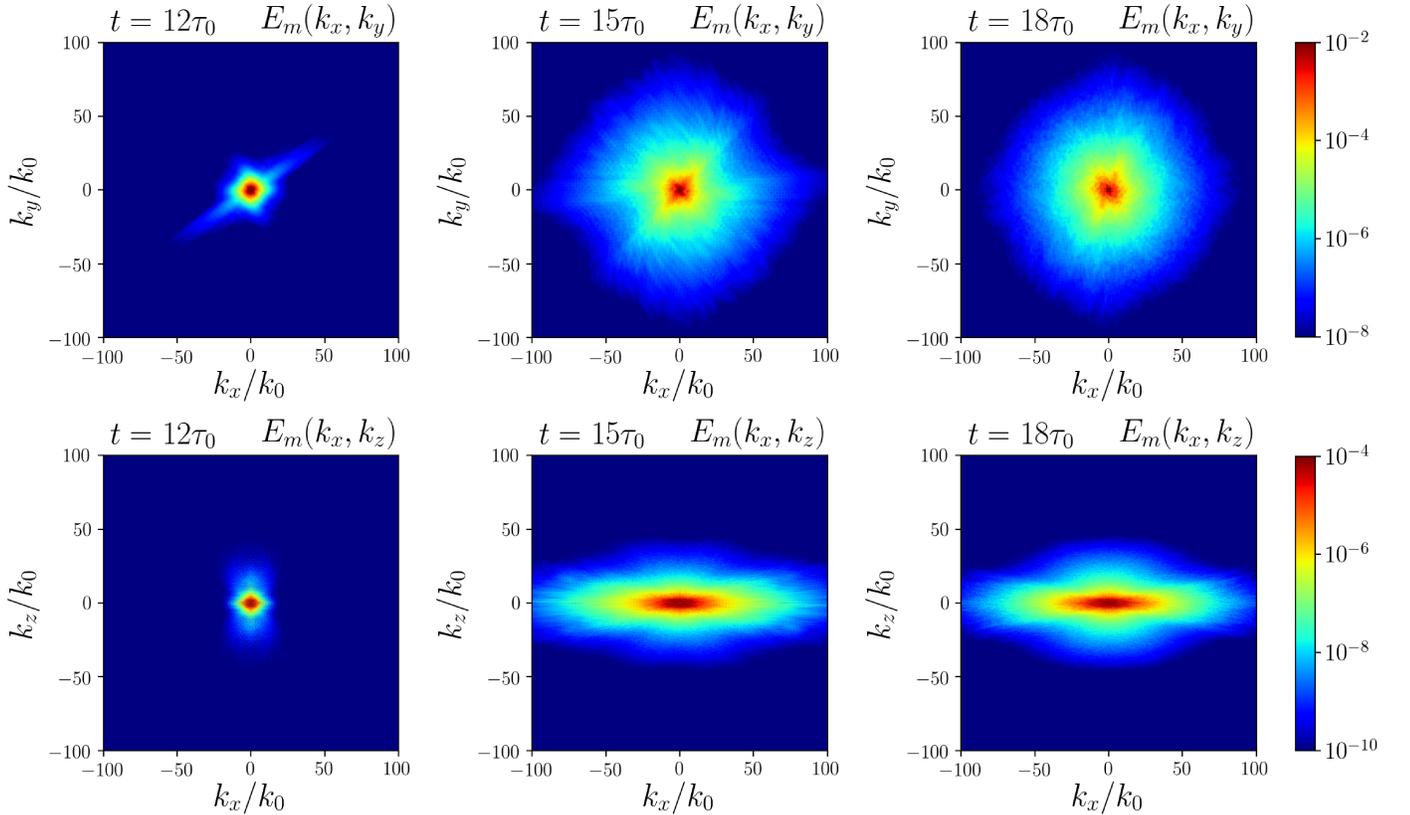} 
	\end{center}
	\vspace{-1em}
	\caption{
		Normalized two-dimensional magnetic energy spectra in the $xy$ directions ($E_m (k_x,k_y)$, upper panels) and in the $xz$ directions ($E_m (k_x,k_z)$, lower panels) from the saturation to nonlinear phase.
		The corresponding times are $t = 12/\tau_0$ (left), $t=15 \tau_0$ (middle), and $t=18 \tau_0$ (right), respectively.
						}
	\vspace{1em}
	\label{fig:powbb}
	\end{figure*}
	
	\begin{figure*}[th]
	\begin{center}
	\includegraphics[width=185mm]{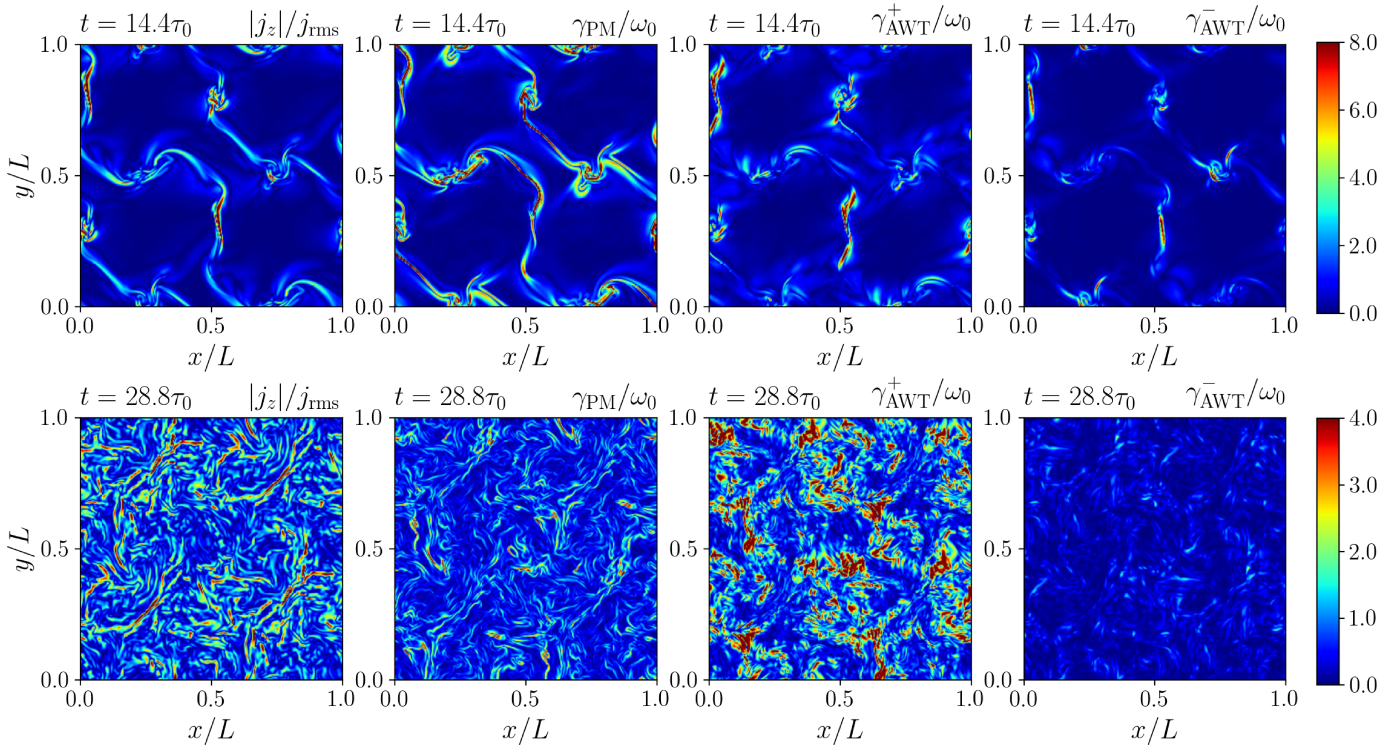} 
	\end{center}
	\vspace{-1em}
	\caption{
		Snapshots of the spatial distributions of $2 \left| j_z \right| / j_{\rm rms}$ (left), $\gamma_{\rm PM} / \omega_0$ (middle-left), $\gamma_{\rm AWT}^+ / \omega_0$ (middle-right), and $\gamma_{\rm AWT}^{-} / \omega_0$ (right).
          	The upper and lower panels correspond to the early nonlinear phase $t = 14.4 \tau_0$ and the late nonlinear phase $t=28.8 \tau_0$, respectively.
						}
	\vspace{1em}
	\label{fig:xy_timescale}
	\end{figure*}

\subsection{Basic equations and solver}
	
	We numerically solve the ideal MHD equations:
	\begin{align}
		&\frac{\partial}{\partial t} \rho + \nabla \cdot \left( \rho \vec{v} \right) =0, \label{eq:mass} \\
  		&\frac{\partial}{\partial t} \left( \rho \vec{v} \right) + \nabla \cdot \left( p_T \hat{\vec{I}} + \rho \vec{v}\vec{v} - \frac{1}{4\pi} \vec{B}\vec{B}  \right) =0, \\
		&\frac{\partial}{\partial t} \vec{B} + \nabla \cdot \left( \vec{v} \vec{B} - \vec{B} \vec{v} \right)  = 0, \\
		&\frac{\partial}{\partial t} e  + \nabla \cdot \left[ \left( e+p_T \right) \vec{v}  - \frac{1}{4 \pi} \vec{B} \left( \vec{v}\cdot\vec{B} \right) \right]  = 0, \\
		&e = \frac{p}{\Gamma-1} + \frac{1}{2} \rho \vec{v}^2 + \frac{\vec{B}^2}{8 \pi}, \ \ \ \ p_T = p + \frac{\vec{B}^2}{8 \pi}. \label{eq:energy}
	\end{align}
	$\Gamma = 1.0001$ is used instead of the adiabatic value $\Gamma = 5/3$ to keep the system almost isothermal.
	Since we are interested in the cascading process, we exclude the diffusion terms; mechanical energy is thermalized by numerical diffusion at the grid scale.
	
	The HLLD Riemann solver \citep{Miyos05} with WENOZ reconstruction \citep{Borge08} and SSP Runge-Kutta method \citep{Shu0088} are used for numerical calculation.
	This is a high-resolution (5th-order-in-space and 3rd-order-in-time) shock-capturing numerical scheme,
	which is appropriate for the simulation of PDI.
	Numerical $\nabla \cdot \vec{B}$ error is removed with the hyperbolic cleaning method \citep{Dedne02}.
	
\subsection{Numerical setup}

	We solve the basic equations  (\ref{eq:mass})-(\ref{eq:energy}) in a rectangular simulation box $[0,L] \times [0,L] \times [0, 3 L]$.
	We apply $288 \times 288 \times 864$ grid points to resolve the computational domain, so that the grid size is isotropic.
	The periodic boundary conditions are used for all the boundaries.
	The initial condition is as follows:
	\begin{align}
		\rho &= \rho_0 + \delta \rho_0, \ p = c_s^2 \rho_0, \\
		v_x  &= \eta v_{A0} \sin \left( k_0  z - \omega_0 t  \right), \  B_x = -  B_0 v_x / v_{A0} \\
		v_y  &= \eta v_{A0} \cos \left( k_0  z - \omega_0 t  \right), \  B_y = - B_0 v_y / v_{A0} \\
		v_z &=0, \ B_z = B_0.
	\end{align}
	Note that the box-averaged density, momentum, magnetic field, and total energy are time-independent because of the periodic boundary.
	$k_0 = 2 \pi / L$ and $\omega_0 = v_{A0} k_0$ are the wave number and frequency of the initial wave, respectively.
	The parent wave period is therefore given as $\tau_0 = 2 \pi / \omega_0 = L / v_{A0}$.
	$c_s$ and $v_{A0} = B_0 / \sqrt{4 \pi \rho_0}$ denote the background sound and Alfv\'en velocities, respectively.
	$\eta$ is the normalized amplitude of the initial wave.
	In this study, we fix $\eta = 0.2$ and $c_s / v_{A0} = 0.2$.
	
	The density fluctuation is imposed to trigger PDI.
	We calculated two cases: three-dimensional (3D) setting and one-dimensional (1D) setting.
	In the 3D setting, the fluctuations are given as
	\begin{align}
		\delta \rho_0 / \rho_0 = \varepsilon \sin \left( k_0 x \right) \sin \left( k_0 y \right) N(z),
		\label{eq:fluctuation}
	\end{align}
	where $\varepsilon \sim 10^{-3}$ is the initial amplitude of density fluctuation and $N(z)$ denotes a random noise function of $z$.
	The 1D setting is defined as
	\begin{align}
		\delta \rho_0 / \rho_0 = \frac{1}{2} \varepsilon N(z).
	\end{align}
	The factor $1/2$ is to make the root-mean-square (rms) value of $\delta \rho_0$ the same as that in the 3D setting.

\section{Volume-averaged quantities  \label{sec:vaq}}	
	
	First, we discuss the time evolution of volume-averaged characteristic quantities: 
	the normalized rms density fluctuation $\delta \rho_{\rm rms} / \rho_0$,  rms angle between current and magnetic field $\phi_{j,B}$, Els\"asser energies $E^{\pm}$ and normalized cross helicity $\sigma_c$.
	$\delta \rho_{\rm rms}$ is defined as
	\begin{align}
		\delta \rho_{\rm rms} = \sqrt{\average{\left( \rho - \rho_0 \right)^2 }} = \left( \rho - \rho_0 \right)_{\rm rms},
	\end{align}
	where $\average{X}$ and $X_{\rm rms}$ denote the volume-average and rms operators, respectively:
	\begin{align}
		\average{X} = \frac{1}{V} \int d\vec{x} X, \ \ \ X_{\rm rms} =  \sqrt{  \frac{1}{V} \int d\vec{x} X^2 },
	\end{align}
	where $V = 3 L^3$ denotes the volume of the simulation box.
	$\phi_{j,B}$ is given by
	\begin{align}
		\phi_{j,B} =  \left[ \arccos^{-1} \left( \frac{\vec{j} \cdot \vec{B}}{\left| \vec{j} \right| \left| \vec{B} \right|} \right) \right]_{\rm rms},
	\end{align}
	where
	\begin{align}
		\vec{j} = \frac{c}{4 \pi} \nabla \times \vec{B}.
	\end{align}
	Normalized Els\"asser energies are defined as:
	\begin{align}
		E^{\pm} = \average{\frac{1}{4} \rho {\vec{\zeta}^{\pm}}^2} / E_0,
	\end{align}
	where $E_0 = \rho_0 v_{A0}^2$ and $\vec{\zeta}^{\pm} = \vec{v}_\perp \mp \vec{B}_\perp/\sqrt{4 \pi \rho}$ are Els\"asser variables \citep{Elsas50}.
	The subscript $\perp$ denotes the components perpendicular to the mean field $B_0 \vec{e}_z$.
	The normalized cross helicity $\sigma_c$ is calculated from $E^{\pm}$ as \citep{DelZa01}
	\begin{align}
		\sigma_c = \frac{E^{+} - E^{-}}{E^{+} + E^{-}}.
	\end{align}
		
	In Figure \ref{fig:fundamental}, we show the time evolution of the volume-averaged quantities defined above by solid lines.
	The corresponding values calculated with the 1D setting are also shown by dashed lines in each panel.
	The blue lines in Figure \ref{fig:fundamental}a are fitted lines that give the growth rates $\gamma$ in the linear phase.

        Figure \ref{fig:fundamental}a shows that the 3D growth rate ($\gamma / \omega_0 = 0.106$) is $26 \%$ smaller than the 1D growth rate ($\gamma / \omega_0 = 0.144$).
        This indicates that the 3D structure of density fluctuation works to reduce the growth rate of PDI.
        Note that both values are smaller than the analytical value ($\gamma / \omega_0 = 0.157$) by \citet{Golds78}.
        $\delta \rho_{\rm rms} / \rho_0$ in the 1D calculation shows an oscillation that is anti-correlated with $E^+$ (Figure \ref{fig:fundamental}a,  \ref{fig:fundamental}c).
        This is a resonant energy exchange between forward Alfv\'en and sound waves \citep{Shoda16}.

        Figure \ref{fig:fundamental}b shows that the alignment between $\vec{j}$ and $\vec{B}$ occurs because of turbulence.
        This is sometimes called selective decay \citep{Biska03}; the magnetic field approaches force-free field turbulence.
        $\phi_{j,B}$ cannot be $0$ because the system has a finite cross helicity \citep{Strib91}.
        
        From Figure \ref{fig:fundamental}c, we can tell the energy dissipation rate of Alfv\'en waves.
        In the later phase of 1D calculation, $E^{-}$ is almost constant because there exists no significant physical mechanism that dissipates backward Alfv\'en waves.
        Meanwhile, in the nonlinear phase of 3D calculation, both $E^{+}$ and $E^{-}$ decreases with time, and the larger component has the smaller decay rate.
        This is the behavior of dynamic alignment \citep{Dobro80,Biska03}; the minor Els\"asser variable decays faster.
	
\section{Anisotropic behavior of turbulence}

        Next, we discuss the structure of turbulence on 2D planes.
        Figure \ref{fig:caby} shows the spatial distributions of the normalized Alfv\'en speed $v_A / v_{A0}$ (where $v_A = B_z / \sqrt{4 \pi \rho}$) on the $xy$ plane (upper panels) 
        and the normalized amplitude $B_y / B_z$ on the $xz$ plane (lower panels) of different $t$:
        from left to right $t=0$ (initial condition), $t = 12\tau_0$ (saturation phase), $t=18 \tau_0$ (early nonlinear phase), and $t=24 \tau_0$ (fully nonlinear phase).
        The inhomogeneity in the upper panel is the trigger of PM, while the filament-like structure is observed in the lower panels because of PM.
        
        As time proceeds, the structure of $B_y / B_z$ becomes aligned with the mean magnetic field $B_0 \vec{e}_z$;
 	the phase structure becomes progressively finer in the direction perpendicular to the mean field.
        Both PM and AWT can generate these structures, and therefore, we cannot distinguish them here.
        PM certainly works in the nonlinear phase because there exists a perpendicular gradient of $v_A$ (upper panels in Figure \ref{fig:caby}); 
        AWT also works because there are bidirectional Alfv\'en waves and the dynamic alignment is observed (Figure \ref{fig:fundamental}c \ref{fig:fundamental}d).
        In Section \ref{sec:pm_versus_awt}, we discuss the dominances of PM and AWT.
 
 	Anisotropy also appears in the energy spectrum.
 	To observe the time evolution of anisotropy, we calculate the normalized magnetic energy spectrum $E_m$ with different times.
	Specifically, we concentrate on the 2D spectrum in the $xy$ directions and $xz$ directions averaged over the other direction, defined as
	\begin{align}
		E_m (k_x,k_y) &= \frac{\Delta k_x \Delta k_y}{3L} \int dz \frac{\vec{B}(k_x,k_y,z)^2}{B_0^2}, \\
		E_m (k_x,k_z) &= \frac{\Delta k_x \Delta k_z}{L} \int dy \frac{\vec{B}(k_x,y,k_z)^2}{B_0^2}, \\
		\Delta k_x &= \Delta k_y = \frac{2 \pi}{L}, \ \ \ \ \Delta k_z = \frac{2 \pi}{3 L}
	\end{align}
	where, for example, $\vec{B} (k_x,y,z)$ denotes the Fourier transformation of $\vec{B} (x,y,z)$ with respect to $x$. 
	
	Figure \ref{fig:powbb} shows $E_m (k_x,k_y)$ (upper panels) and $E_m (k_x,k_z)$ (lower panels) at $t=12 \tau_0$ (left), $t=15 \tau_0$ (middle), and $t=18 \tau_0$ (right), respectively.
        The upper panels show isotropic behavior, indicating that the turbulence is axisymmetric with respect to the mean magnetic field.
        In contrast, the lower panels show anisotropic distributions; the contour is elongated along $k_x$ in the nonlinear phase.
        This elongation shows that the cascading proceeds faster in the perpendicular ($x$) direction than in the parallel ($z$) direction, which is consistent with the classical theory of AWT \citep{Goldr95}.
        In fact, the typical aspect ratio ($k_z / k_x$) of the spectrum is approximately the same as the nonlinearity ($\delta v / v_A$), and this indicates the critical balance of AWT \citep{Goldr95}.

\section{Phase mixing versus Alfv\'en wave turbulence \label{sec:pm_versus_awt}}
		
	Both PM (phase mixing) and AWT (Alfv\'en wave turbulence) yield anisotropy, and this similarity makes it difficult to distinguish the physical processes.
	Here, by calculating the timescale of each process, we aim to clarify which process is dominant in a certain phase.
	
	To estimate the timescale of PM and AWT, we derive the analytical expression of the timescale of PM and AWT.
	Because both processes are caused by Alfv\'en waves, reduced MHD (RMHD) approximation is convenient.
	In the absence of parallel flow and parallel gradients of density and magnetic field, the Alfv\'en wave propagation is described as follows \citep{Verdi07}:
	\begin{align}
		\left[ \partial / \partial t \pm v_{A \parallel} \nabla_\parallel \right] \vec{\zeta}^\pm =  - \left( \vec{\zeta}^\mp \cdot \nabla_\perp \right) \vec{\zeta}^\pm, \label{eq:vv07}
	\end{align}
	where
	\begin{align}
		\nabla_\parallel = \vec{e}_z \frac{\partial}{\partial z}, \ \ \ \ \nabla_\perp =  \vec{e}_x \frac{\partial}{\partial x} + \vec{e}_y \frac{\partial}{\partial y}.
	\end{align}
	The second term on the left hand side is the term for propagation and PM, 
	while the first term on the right hand side corresponds to AWT.
	Even though the RMHD approximation does not hold in our simulation, Eq. (\ref{eq:vv07}) approximately describes the Alfv\'en wave propagation and provides a general understanding of the physical process that is not related to compressibility.
	
	We begin by deriving the timescale of PM.
	Nonlinear terms are ignored here because PM is a linear process triggered by the perpendicular variance of the Alfv\'en speed.
	Taking the perpendicular rotation ($\nabla_\perp \times$) of Eq. (\ref{eq:vv07}), we obtain
	\begin{align}
		\left[ \partial / \partial t \pm v_{A \parallel} \nabla_\parallel \right] \left( \nabla_\perp \times \vec{\zeta}^\pm \right) =  \mp \left( \nabla_\perp  v_{A \parallel} \right) \times \frac{\partial}{\partial z} \vec{\zeta}^{\pm}.
 	\end{align}
	The right-hand-side term represents the phase mixing.
	From this equation, the growth rate of PM ($\gamma_{\rm PM}$) is approximately given as follows:
	\begin{align}
		\gamma_{\rm PM} \approx \left| \nabla_\perp v_{A \parallel} \right|.
	\end{align}
	
	Next, we calculate the timescale of AWT.
	Ignoring the linear term in Eq. (\ref{eq:vv07}), the decay rate of Alfv\'en wave turbulence is given as
	\begin{align}
		\gamma^{\pm}_{\rm AWT} \approx \frac{ \left| \left( \tilde{\vec{\zeta}}^\mp \cdot \nabla_\perp \right) \vec{\zeta}^\pm \right| }{\average{\vec{\zeta}^\pm}_{\rm rms}},
	\end{align}
	where $\tilde{\vec{\zeta}}^\mp$ denote the fluctuating parts of Els\"asser variables:
	\begin{align}
		\tilde{\vec{\zeta}}^\mp = \vec{\zeta}^\mp - \frac{1}{L^2} \int dx dy \vec{\zeta}^\mp.
	\end{align}
	We evaluate the nonlinear operator as  $\tilde{\vec{\zeta}}^\mp \cdot \nabla_\perp$ instead of $\vec{\zeta}^\mp \cdot \nabla_\perp$ because the perpendicularly uniform mode does not contribute to the wave distortion (energy cascading).

	In Figure \ref{fig:xy_timescale}, we show the spatial distribution of $j_z / j_{\rm rms}$, $\gamma_{\rm PM}$ and $\gamma_{\rm AWT}^{\pm}$ on the $xy$ plane in an early nonlinear phase ($t = 14.4 \tau_0$, upper panels) and a late nonlinear phase ($t = 28.8 \tau_0$, lower panels).
	Note that $j_z / j_{\rm rms}$ indicates the degree of development of the perpendicular cascading.
	In an early phase, $\gamma_{\rm PM}$ is larger than $\gamma_{\rm AWT}^{\pm}$ and it is spatially correlated with $j_z / j_{\rm rms}$ and $\gamma_{\rm AWT}^{\pm}$.
	This shows that the early nonlinear phase is dominated by PM-driven turbulence \citep{Magya17}.
	In the later phase, the magnitude relation is reversed; $\gamma_{\rm AWT}^{+}$ becomes the largest and this indicates that the later phase is characterized by Alfv\'en wave turbulence.
	Specifically, AWT in the later phase is imbalanced ($E^+ \ll E^-$, Figure \ref{fig:fundamental}c), and thus the dynamic alignment proceeds (Figure \ref{fig:fundamental}d).
		
	\begin{figure}[t!]
	\begin{center}
	\includegraphics[width=80mm]{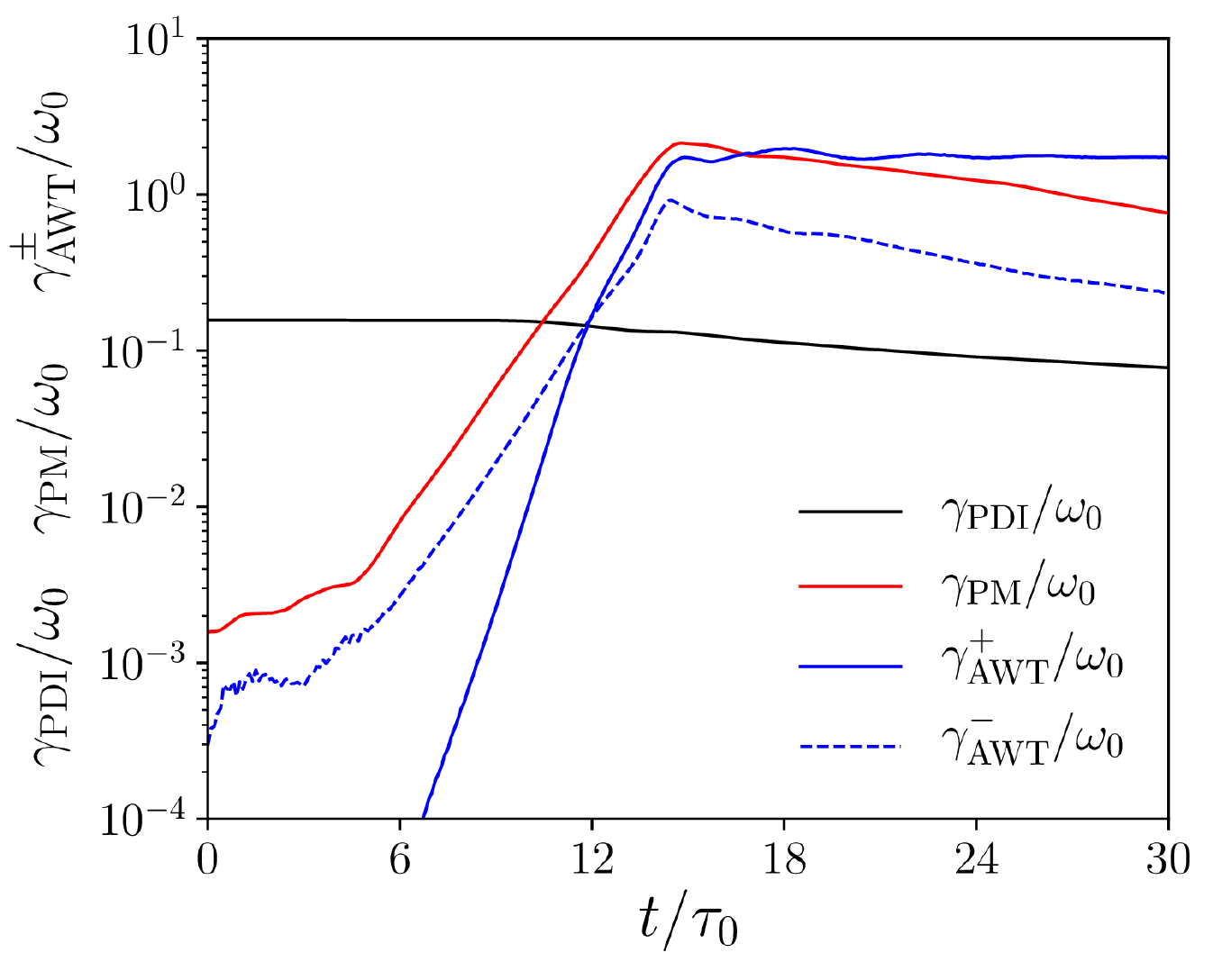} 
	\end{center}
	\vspace{-1em}
	\caption{
			Time evolution of rms timescales of phase mixing $\gamma_{\rm PM}$ (red solid), Alfv\'en wave turbulence of forward propagating mode $\gamma_{\rm AWT}^+$ (blue solid) and backward propagating mode $\gamma_{\rm AWT}^-$ (blue dashed), 
			and parametric decay instability $\gamma_{\rm PDI}$ normalized by the initial-wave angular frequency $\omega_0$.
						}
	\vspace{1em}
	\label{fig:timescale}
	\end{figure}
	
	Figure \ref{fig:timescale} shows the rms values of $\gamma_{\rm PM}$ (red solid line), $\gamma_{\rm AWT}^{+}$ (blue solid line), and $\gamma_{\rm AWT}^{-}$ (blue dashed line) versus time.
	In addition to these, we also calculate the normalized growth rate of PDI ($\gamma_{\rm PDI}$) from the dispersion relation given by \citet{Golds78} and show its time evolution with a black line.
	
	Figure \ref{fig:timescale} gives some important indications.
	First, the system is dominated by different processes, depending on the phase.
	The initial phase ($0 \leq t / \tau_0 \lesssim 8$) is dominated by the growth of PDI because $\gamma_{\rm PDI}$ is the largest.
	Before the saturation to early nonlinear phase ($8 \lesssim t / \tau_0 \lesssim 16$), PM becomes active because of the large density fluctuation (Figure \ref{fig:fundamental}a).
	Finally, in the fully nonlinear phase ($16 \lesssim t / \tau_0$), the system is characterized by imbalanced ($\gamma_{\rm AWT}^{+} \gg \gamma_{\rm AWT}^{-}$) AWT.
	 
	Second, because the timescale of $\gamma_{\rm PM}$ is clearly correlated with $\delta \rho_{\rm rms} / \rho_0$, 
	PM should be of importance in the large-density-fluctuation regions.
	In the corona and solar wind, such a region is either the coronal bottom \citep{Raymo14} or the wind acceleration region \citep{Miyam14}.
	Specifically, the fast saturation of nonthermal line broadening in the corona \citep{Hahn013} possibly comes from PM because of the presence of large density fluctuations near the coronal bottom \citep{Raymo14}.
		
	M.S. is supported by the Leading Graduate Course for Frontiers of Mathematical Sciences and Physics (FMSP) and Grant-in-Aid for Japan Society for the Promotion of Science (JSPS) Fellows.
	T.Y. is supported by JSPS KAKENHI Grant Number 15H03640.
	Numerical computations were carried out on Cray XC30 at Center for Computational Astrophysics, National Astronomical Observatory of Japan.

\end{document}